\documentclass[12pt]{article}
\usepackage{amsmath,amssymb,cite}
\advance\hoffset by -1.1truecm
\advance\voffset by -1.0truecm
\newcommand{\be}{\begin{equation}}
\newcommand{\ee}{\end{equation}}
\newcommand{\bea}{\begin{eqnarray}}
\newcommand{\eea}{\end{eqnarray}}
\newcommand{\pp}{~.}
\newcommand{\vv}{~,}
\newcommand{\cc}{{\mathbb{C}}}
\newcommand{\rr}{{\mathbb{R}}}

\numberwithin{equation}{section}

\begin{document}

\title{\begin{flushright}
 \small SU-4252-813\\IISc/CHEP/7/05\\DSF-20/05
\end{flushright} \vspace{0.25cm} Spin and Statistics on the
 Groenewold-Moyal Plane:Pauli-Forbidden Levels and Transitions}
\author{A.P. Balachandran$^a$\footnote{bal@phy.syr.edu},
G. Mangano$^{a,b}$\footnote{mangano@na.infn.it}, A.
Pinzul$^a$\footnote{apinzul@phy.syr.edu}
\, and S. Vaidya$^c$\footnote{vaidya@cts.iisc.ernet.in}\\ \\
$^a$ \begin{small} Department of Physics, Syracuse University, Syracuse NY,
13244-1130, USA. \end{small} \\ $^b$ \begin{small} INFN, Sezione di Napoli
and Dipartimento di Scienze Fisiche,
\end{small} \\ \begin{small}
Universit\`{a} di Napoli {\it Federico II}, Via Cintia, I-80126 Napoli, Italy.
\end{small} \\ $^c$ \begin{small} Centre for High Energy Physics, Indian
Institute of Science, \end{small} \\ \begin{small} Bangalore, 560012, India.
\end{small}}
\date{\empty}

\maketitle
\begin{abstract}
The Groenewold-Moyal plane is the algebra ${\cal A}_\theta(\rr^{d+1})$
of functions on $\rr^{d+1}$ with the $\ast$-product as the
multiplication law, and the commutator $\left[ \hat{x}_\mu,\hat{x}_\nu
\right] =i \theta_{\mu \nu}$ ($\mu,\nu=0,1,...,d$) between the
coordinate functions.  Chaichian et al. \cite{Chaichian:2004za} and
Aschieri et al. \cite{aschieri} have proved that the Poincar\'{e}
group acts as automorphisms on ${\cal A}_ \theta({\rr^{d+1}})$ if the
coproduct is deformed. (See also the prior work of Majid \cite{majid}, Oeckl
\cite{Oeckl:2000eg} and Grosse et al \cite{gms}). In fact, the
diffeomorphism group with a deformed coproduct also does so according
to the results of \cite{aschieri}. In this paper we show that for this
new action, the Bose and Fermi commutation relations are deformed as
well. Their potential applications to the quantum Hall effect are
pointed out. Very striking consequences of these deformations are the
occurrence of Pauli-forbidden energy levels and transitions. Such new
effects are discussed in simple cases. \\ \\ PACS numbers: 11.10.Nx,
11.30.Cp
\end{abstract}

{\it Dedicated to Rafael Sorkin, our friend and  teacher, and a true
  and creative seeker of knowledge.} 

\section{Introduction}
\label{sec:intro} The Groenewold-Moyal plane is the algebra ${\cal
A}_\theta(\rr^{d+1})$ of functions on $\rr^{d+1}$ with the
$\ast$-product $\alpha \ast_\theta \beta$ between functions
$\alpha$ and $\beta$ as the product law, where
\begin{eqnarray}
\alpha \ast_\theta \beta \;(x) &=& \left[ \alpha \exp\left(\frac{i}{2}
\overleftarrow{\partial_\mu} \theta^{\mu \nu}
\overrightarrow{\partial_\nu} \right) \beta \right] (x) \vv
\label{starprod} \\
\theta^{\mu \nu}&=&-\theta^{\nu \mu} \in \rr \vv \   x=
(x^0,x^1,\ldots,x^d) \pp \nonumber
\end{eqnarray}

The Poincar\'{e} group $\mathcal{P}$ acts on $\rr^{d+1}$ and hence
on its smooth functions $C^{\infty}(\rr^{d+1})$ regarded just as a
vector space. If $g \in \mathcal{P}$ and $g: x \rightarrow g x$,
then for $\gamma \in C^{\infty}(\rr^{d+1})$
\begin{equation}\label{gaction}
( g \gamma) (x) = \gamma (g^{-1} x) \pp
\end{equation}
However, in general
\begin{equation}
(g \alpha) \ast_\theta (g \beta) \neq g (\alpha \ast_\theta \beta)
\label{gastarb} \vv
\end{equation}
so that this action of $\mathcal{P}$ is not an automorphism of
${\cal A}_\theta(\rr^{d+1})$.

Similar remarks can be made generically about any group which acts on
$\rr^{d+1}$ and in particular about the diffeomorphism group
$\mathcal{D}$. Only a limited group of transformations, such as
translations, gives the equality in (\ref{gastarb}). Nevertheless,
there is a way to avoid this limitation with $\mathcal{D}$. It
involves introducing a new deformed coproduct $\Delta_\theta$ on
$\mathcal{D}$. The revival of this idea in recent times is due to
\cite{Chaichian:2004za,aschieri,Dimitrijevic:2004rf,Oeckl:2000eg}.
But its origins can be traced back to Drin'feld \cite{drinfeld} in
mathematics. This Drin'feld twist leads naturally to deformed
$R$-matrices and statistics for quantum groups, as discussed by Majid
\cite{majid}. Subsequently, Fiore and Schupp \cite{fiore1} and Watts
\cite{watts1,watts2} explored the significance of the Drin'feld twist
and $R$-matrices while Fiore \cite{fioresolo1, fioresolo2} and Fiore
and Schupp \cite{fiore2}, Oeckl \cite{Oeckl:2000eg} and Grosse et al
\cite{gms} studied the importance of $R$-matrices for
statistics. Oeckl \cite{Oeckl:2000eg} and Grosse et al \cite{gms} also
developed quantum field theories using different and apparently
inequivalent approaches, the first on the Moyal plane and the second
on the $q$-deformed fuzzy sphere. Recent work, including ours, has
significant overlap with the earlier literature. We share many
features in particular with \cite{gms,Oeckl:2000eg}.

In \cite{aschieri,Dimitrijevic:2004rf} the authors focused on
$\mathcal{D}$ and developed Riemannian geometry and gravity theories
based on $\Delta_\theta$, while \cite{Chaichian:2004za} focused on the
Poincar\'{e} subgroup $\mathcal{P}$ of $\mathcal{D}$ and explored the
consequences of $\Delta_\theta$ for quantum field theories. Twisted
conformal symmetry was discussed by \cite{matlock}. We explain the
basics of all this work in Section 2.

In Section 3, we discuss the impact of the deformed tensor product on
the Bose and Fermi commutation relations. In fact, they are also
deformed. We give an explicit formula for the new
creation-annihilation operators in terms of the standard ($\theta^{\mu
\nu}=0$) ones. State vectors can still be classified by the
irreducible representations of the permutation group, but the action
of the latter on the Hilbert space is deformed as well.

Previous research on the spin-statistics theorem on ${\cal
A}_\theta(\rr^{d+1})$ is due to Alvarez-Gaum\'{e} and Vazquez-Mozo
\cite{Alvarez-Gaume:2003mb}, but they do not use the deformed
coproduct on $\mathcal{P}$.

In Section 4, we construct the second quantization formalism
corresponding to the deformed commutation relations, introducing also
the corresponding symmetry under permutations of physical states.

In Section 5, we argue that excitations in the quantum Hall effect
should be described by deformed statistics.

Finally, in Section 6, we discuss the possible phenomenological
implications of the deformed commutation relations, considering in
particular the case of systems of fermionic identical particles.  We
show that there exist state vectors of the system which violate the
Pauli exclusion principle. There are quite stringent tests on Pauli
violating transitions in nuclear (see for example \cite{sk,borexino}
and references therein) and atomic systems \cite{xrays}, and crystals
\cite{greenberg}, so that the energy scale associated with
$\theta^{\mu \nu}$ (whose dimension is inverse squared energy) can be
severely constrained. This issue will be studied in more detail later.

In another work \cite{bpq}, it is proved that UV-IR mixing is entirely
absent for quantum field theories on ${\cal A}_\theta ({\mathbb
R}^{d+1})$ with the deformed statistics.

\section{The Deformed Coproduct}

\subsection{\it{Tensor Product of Representations}}

Suppose that a group $G$ acts on a complex vector space $V$ by a
representation $\rho$. We denote this action by
\begin{equation}
v \rightarrow \rho(g) v \vv
\label{rhov}
\end{equation}
for $g \in G$ and $v \in V$. Then the group algebra $G^*$ also acts on
$V$.  A typical element of $G^*$ is
\begin{equation}
\int dg \,\alpha(g)\, g, \,\,\,\,\, \alpha(g) \in \cc \vv
\end{equation}
where $dg$ is a measure on $G$. Its action is
\begin{equation}
v \rightarrow \int dg \,\alpha(g) \, \rho (g) \, v \pp
\label{actgstar}
\end{equation}

Both $G$ and $G^*$ act on $V \otimes_\cc V$, the tensor product of
$V$'s over $\cc$, as well. These actions are usually taken to be
\begin{equation}
v_1 \otimes v_2 \rightarrow \left[ \rho(g) \otimes \rho(g) \right]
(v_1 \otimes v_2 ) = \rho(g) v_1 \otimes \rho(g) v_2 \vv
\label{acttens}
\end{equation}
and
\begin{equation}
v_1 \otimes v_2 \rightarrow
\int dg \, \alpha(g) \, \rho(g) v_1 \otimes \rho(g) v_2
\end{equation}
respectively, for $v_1, v_2 \in V$.

In Hopf algebra theory, the action of $G$ and $G^*$ on tensor products
is formalized using the coproduct $\Delta$, a homomorphism from $G^*$
to $G^* \otimes G^*$, which on restriction to $G$ gives a homomorphism
from $G$ to $G^* \otimes G^*$. This restriction specifies $\Delta$ on
all of $G^*$ by linearity. Thus if
\begin{eqnarray}
&& \Delta: \, g \rightarrow \Delta(g) \vv \\
&& \Delta(g_1)\Delta(g_2)=\Delta(g_1g_2) \vv
\end{eqnarray}
we have
\begin{equation}
\Delta \left(\int dg \, \alpha(g) \, g \right) = \int dg \, \alpha(g)
\, \Delta(g) \pp
\end{equation}

For the familiar choice $\Delta(g) = g \otimes g$, the action
(\ref{acttens}) can be written as
\begin{equation}
v_1 \otimes v_2 \rightarrow \left[\rho \otimes \rho \right]
\Delta(g) v_1 \otimes v_2 \pp
\label{coprod}
\end{equation}
But any choice of coproduct will do to define an action of $G$ on $V
\otimes V$ using (\ref{coprod}).

Likewise, if $G$ acts on vector spaces $V$ and $W$ by representations
$\rho$ and $\sigma$, respectively, and $\Delta$ is a coproduct on $G$,
$G$ can act on $V \otimes W$ according to
\begin{equation}
v \otimes w \rightarrow \left[\rho \otimes \sigma \right] \Delta(g) v
\otimes w \vv
\end{equation}
for $v \in V$, $w \in W$. This action extends by linearity to an
action of $G^*$.

Not all choices of $\Delta$ are equivalent. In particular the
irreducible representations (IRR's), which can occur in the reduction
of $\left[ \rho \otimes \sigma \right]$ can depend upon
$\Delta$. Examples of this sort perhaps occur for the Poincar\'{e}
group.

\subsection{\it{The Carrier of Group Action is an Algebra}}

Until now we assumed only that $V,W$ are vector spaces. Suppose next
that $V$ is an algebra ${\cal A}$ (over $\cc$). In that case, as
discussed by \cite{Chaichian:2004za,aschieri} there is also a
compatibility condition on $\Delta$. It comes about as follows.

As ${\cal A}$ is an algebra, we have a rule for taking products of
elements of ${\cal A}$. That means that there is a multiplication map
\begin{eqnarray}
m: {\cal A} \otimes {\cal A} \rightarrow {\cal A} \vv \\
\alpha \otimes \beta
\rightarrow m (\alpha \otimes \beta) \vv \nonumber \label{multmap}
\end{eqnarray}
for $\alpha,\beta \in {\cal A}$, the product $\alpha \beta$ being $m
(\alpha \otimes \beta)$.

It is now essential that $\Delta$ be compatible with $m$. That means
that if we transform $\alpha \otimes \beta$ by $g$-action and then
apply $m$, it should be equal to the $g$-transform of $m (\alpha
\otimes \beta)$:
\begin{equation}
m \left( (\rho \otimes \rho) \Delta(g) \left(\alpha \otimes \beta
\right)\right)=
\rho(g) m ( \alpha \otimes \beta) \pp
\label{compatib}
\end{equation}
This result is encoded in the commutative diagram
\begin{equation}
\begin{array}{ccc}\label{diag}
\alpha \otimes \beta & \longrightarrow & ( \rho \otimes \rho ) \Delta
(g) \alpha \otimes \beta \\
& &  \\ m \,\, \downarrow  & & \downarrow \,\, m \\ &  & \\ m(\alpha
\otimes
\beta) &
\longrightarrow & \rho(g) m (\alpha \otimes \beta)
\end{array}
\end{equation}
If such a $\Delta$ can be found, $G$ is an automorphism of ${\cal
A}$. In the absence of such a $\Delta$, $G$ does not act on ${\cal
A}$.

\subsection{\it{The Case of the Groenewold-Moyal Plane}}

In the Groenewold-Moyal plane, the multiplication map depends on
$\theta^{\mu \nu}$ and will be denoted by $m_\theta$. It is defined by
\begin{equation}
m_\theta ( \alpha \otimes \beta) = m_0 \left( e^{-\frac{i}{2} (-i
\partial_\mu) \theta^{\mu \nu} \otimes (-i \partial_\nu) } \alpha
\otimes \beta \right) \vv
\label{multmoyal}
\end{equation}
where $m_0$ is the point-wise multiplication
\begin{equation}
m_0 ( \gamma \otimes \delta) (x) := \gamma(x) \delta (x)
\end{equation}
of any two functions $\gamma$ and $\delta$.

We introduce the notation
\begin{equation}
F_\theta = e^{-\frac{i}{2} (-i \partial_\mu) \theta^{\mu \nu}
\otimes (-i \partial_\nu) }
\label{ftheta}
\end{equation}
for the factor appearing in (\ref{multmoyal}) so that
\begin{equation}
m_\theta ( \alpha \otimes \beta) = m_0 \left( F_\theta \alpha \otimes \beta
\right) \pp
\label{multmoyal2}
\end{equation}

Let $g\in \mathcal{D}$ act on $\rr^{d+1}$ by $x \rightarrow g x$ and
hence on functions by $\alpha \rightarrow \rho(g) \alpha$ where the
representation $\rho$ is canonical:
\begin{equation}
(\rho(g) \alpha) (x) = \alpha ( g^{-1} x) \pp
\label{canonrepr}
\end{equation}
(This action was denoted in Eq.(\ref{gaction}) omitting the symbol
$\rho$.) The important observation is that it can act on ${\cal
A}_\theta(\rr^{d+1}) \otimes {\cal A}_\theta(\rr^{d+1})$ as well
compatibly with $m_\theta$ if a new coproduct $\Delta_\theta$ is used,
where
\begin{equation}
\Delta_\theta (g) =e^{\frac{i}{2} P_\mu {\otimes} \theta^{\mu \nu} P_\nu }
(g {\otimes} g) e^{-\frac{i}{2} P_\mu {\otimes} \theta^{\mu \nu} P_\nu
} = \hat{F}^{-1}_\theta (g \otimes g) \hat{F}_\theta \vv
\label{newcoprod}
\end{equation}
$P_\mu$ being the generators of translations. On functions, that is,
in the representation $\rho$, it becomes $-i \partial_\mu$, so that
the two factors in (\ref{newcoprod}) can be expressed in terms of
$F_\theta$ and its inverse.

We can check that $\Delta_\theta$ is compatible with $m_\theta$ as follows
\begin{eqnarray}
m_\theta \left( (\rho \otimes \rho) \Delta_\theta(g) ( \alpha \otimes
\beta ) \right) &=&
m_0 \left( F_\theta (F_\theta^{-1} \rho(g) \otimes
\rho(g) F_\theta) \alpha \otimes \beta \right) \nonumber \\
&=& \rho(g) \left( \alpha \ast_\theta \beta \right), \quad
 \alpha,\beta \in {\cal A}_\theta(\rr^{d+1})
\label{proofcomp}
\end{eqnarray}
as required.

The action of the Poincar\'{e} group on tensor products of plane waves
is simple. For the momentum $p=(p_0, p_1,...p_d) \in \rr^{d+1}$, let
${\bf e}_p \in {\cal A}_\theta(\rr^{d+1})$ where
\begin{equation}
{\bf e}_p(x) = e^{i p {\cdot} x}, \quad p {\cdot} x = p_\mu x^\mu \pp
\label{plane}
\end{equation}

In the case of the Poincar\'{e} group, if $\exp( i P {\cdot} a) $ is a
translation,
\begin{equation}
(\rho \otimes \rho)\Delta_\theta \left( e^{i P {\cdot} a} \right)
{\bf e}_p \otimes {\bf e}_q = e^{i (p+q) {\cdot} a}  {\bf e}_p \otimes
{\bf e}_q  \vv
\label{planetrans}
\end{equation}
while if $\Lambda$ is a Lorentz transformation
\begin{equation}
(\rho \otimes \rho)\Delta_\theta(\Lambda) {\bf e}_p \otimes {\bf e}_q
= \left[e^{\frac{i}{2}(\Lambda p)_\mu \theta^{\mu \nu}
    (\Lambda q)_\nu } e^{-\frac{i}{2} p_\mu \theta^{\mu \nu}
    q_\nu } \right] {\bf e}_{\Lambda p} \otimes {\bf e}_{\Lambda q} \pp
\label{planelorentz}
\end{equation}
Thus the coproduct on translations is not affected while the coproduct
for the Lorentz group is changed.

\subsection{\it{Action on Fourier Coefficients}}

If $\varphi$ is a scalar field, we can regard it as an element of
${\cal A}_\theta(\rr^{d+1})$. If its Fourier representation is
\begin{equation}
\varphi = \int d \mu(p) \, \tilde{\varphi}(p) \, {\bf e}_p \vv
\label{field}
\end{equation}
where $d \mu(p)$ is a Lorentz-invariant measure, then
\begin{eqnarray}
\rho(\Lambda) \varphi &=& \int d \mu(p) \, \tilde{\varphi}(p) \, {\bf
e}_{\Lambda p} = \int d \mu(p) \, \tilde{\varphi}(\Lambda^{-1} p) \, {\bf
e}_{p} \vv \\
\rho \left( e^{i P {\cdot} a} \right)  \varphi &=& \int d \mu(p) \,
e^{i p {\cdot} a}\tilde{\varphi}(p) \, {\bf e}_{p} \pp
\end{eqnarray}
Thus the representation $\tilde{\rho}$ of the Poincar\'{e} group on
$\tilde{\varphi}$ is specified by
\begin{eqnarray}
\left(\tilde{\rho}(\Lambda) \tilde{\varphi} \right) (p) &=&
\tilde{\varphi}(\Lambda^{-1} p) \vv \\
\left(\tilde{\rho}\left( e^{i P {\cdot} a} \right) \tilde{\varphi} \right) (p)
&=& e^{i p {\cdot} a} \tilde{\varphi}(p) \pp
\end{eqnarray}

If $\chi$ is another field of ${\cal A}_\theta(\rr^{d+1})$,
\begin{equation}
\chi = \int d \mu(p) \, \tilde{\chi}(p) \, {\bf e}_p \vv
\label{field2}
\end{equation}
then
\begin{equation}
\varphi \otimes \chi = \int d \mu(p) \, d \mu(q) \, \tilde{\varphi}(p)
\tilde{\chi}(q) \, {\bf e}_{p} \otimes {\bf e}_{q} \label{phitimeschi}
\pp
\end{equation}
Using (\ref{planetrans}), we see that the action of translations on
$\tilde{\varphi} \otimes \tilde{\chi}$ is
\begin{equation}
\Delta_\theta \left( e^{i P {\cdot} a} \right) \left( \tilde{\varphi} \otimes
\tilde{\chi} \right) (p,q) = e^{i (p+q) {\cdot} a} \tilde{\varphi}(p)
\tilde{\chi}(q)
\label{reprphichi1} \pp
\end{equation}
Using (\ref{reprphichi1}) we can deduce the action of twisted Lorentz
transformations to be
\begin{equation}
\Delta_\theta(\Lambda) \left( \tilde{\varphi} \otimes \tilde{\chi}
\right) (p,q) =  \tilde{F}^{-1}_\theta\left( \Lambda^{-1}p,
\Lambda^{-1}q \right) \tilde{F}_\theta\left( p, q \right)
\tilde{\varphi}(\Lambda^{-1}p)  \tilde{\chi}(\Lambda^{-1}q)
\label{reprphichi2} \pp
\end{equation}
Here
\begin{equation}
\tilde{F}_\theta\left( r, s \right) := e^{-\frac{i}{2} r_\mu
  \theta^{\mu \nu} s_\nu }
\end{equation}
and we have omitted writing $\rho \otimes \rho$ in front of
$\Delta_\theta$'s.

We remark that had we used (\ref{planelorentz}) to deduce the
transformation law for the Fourier coefficients, we would have got
$\tilde{F}_\theta\left( \Lambda^{-1}p, \Lambda^{-1}q \right)
\tilde{F}^{-1}_\theta\left( p, q \right) \tilde{\varphi}
(\Lambda^{-1}p) \tilde{\chi}(\Lambda^{-1}q)$ for the right-hand side
of (\ref{reprphichi2}). We will use (\ref{reprphichi2}) hereafter as
it can be deduced from the conventional action of $P_\mu$ given by
(\ref{reprphichi1}).

\section{Quantum Fields and Spin-Statistics}

A free relativistic scalar quantum field $\varphi$ of mass $m$ can be
expanded as
\begin{equation}
\varphi = \int \frac{d^d p}{2 p_0} \left( a(p) \,{\bf e}_p + a^\dagger
(p) {\bf e}_{-p} \right)
\vv \label{phiexpans}
\end{equation}
where $p_0=\sqrt{\left|\vec{p}\right|^2 +m^2} \geq 0$, and $a(p),
a^\dagger(p)$ are subject to suitable relations to be stated below. If
$c(p), c^\dagger(p)$ are the limits of these operators when
$\theta^{\mu \nu}=0$, these relations are
\begin{eqnarray}
\left[ c(p), c(q) \right] & = &  \left[ c^\dagger(p),
c^\dagger(q) \right] = 0 \vv \label{standcomm1} \\
\left[ c(p), c^\dagger(q) \right] & = & 2 p_0 \delta^d(p-q) \pp
\label{standcomm2}
\end{eqnarray}
We now argue that such relations are incompatible for $\theta^{\mu
\nu} \neq 0$. Rather $a(p)$ and $a^\dagger(p)$ fulfill certain
deformed relations which reduce to (\ref{standcomm1}),
(\ref{standcomm2}) for $\theta^{\mu \nu} = 0$. We may therefore say
that statistics is deformed, though this is not entirely precise, as
we discuss later.

Similar deformations occur for the operator relations of all tensorial
and spinorial quantum fields.

Suppose now that
\begin{equation}
a(p) a(q) = \tilde{G}_\theta(p,q) a(q) a(p) \vv
\label{newcomm1}
\end{equation}
where $\tilde{G}_\theta$ is a $\cc$-valued function of $p$ and $q$ yet
to be determined. In particular, if $U(\Lambda)$ and $U(\exp (i P
{\cdot} a))$ are the operators implementing the Lorentz
transformations and translations respectively on the quantum Hilbert
space,
\begin{eqnarray}
U(\Lambda) \tilde{G}_\theta(p,q) U(\Lambda)^{-1} &=&
\tilde{G}_\theta(p,q) \vv \\
U(\exp (i P {\cdot} a)) \tilde{G}_\theta(p,q) U(\exp (i P {\cdot}
a))^{-1} &=& \tilde{G}_\theta(p,q) \pp
\end{eqnarray}

The transformations of $a(p) a(q) = (a \otimes a) (p,q)$ and $a(q)
a(p)$ are determined by $\Delta_\theta$. Hence conjugating
(\ref{newcomm1}) by $U(\Lambda)$, we get
\begin{eqnarray}
&& \tilde{F}^{-1}_\theta\left( \Lambda^{-1}p, \Lambda^{-1}q \right)
\tilde{F}_\theta\left( p, q \right) a(\Lambda^{-1}p) a(\Lambda^{-1}q)
= \nonumber \\
&& =  \tilde{G}_\theta(p,q) \tilde{F}^{-1}_\theta\left( \Lambda^{-1}q,
\Lambda^{-1}p \right) \tilde{F}_\theta\left( q, p \right)
a(\Lambda^{-1}q) a(\Lambda^{-1}p) \vv
\end{eqnarray}
or, on using $\tilde{F}_\theta\left( r, s \right)=
\tilde{F}^{-1}_\theta\left( s,r \right)$,
\begin{equation}
a(\Lambda^{-1}p) a(\Lambda^{-1}q)= \tilde{G}_\theta(p,q)
\tilde{F}^{-2}_\theta\left( \Lambda^{-1}q, \Lambda^{-1}p \right)
\tilde{F}^{2}_\theta\left( q, p \right) a(\Lambda^{-1}q)
a(\Lambda^{-1}p) \pp
\label{constraint1}
\end{equation}
Using (\ref{newcomm1}) after changing $p$ to $\Lambda^{-1}p$ and $q$
to $\Lambda^{-1}q$, we get
\begin{equation}
\tilde{G}_\theta(\Lambda^{-1} p, \Lambda^{-1} q)
\tilde{F}^{2}_\theta\left(\Lambda^{-1} q, \Lambda^{-1} p \right)
= \tilde{G}_\theta(p,q) \tilde{F}^{2}_\theta\left( q,p \right) \vv
\label{conG}
\end{equation}
whose solution is
\begin{equation}
\tilde{G}_\theta(p,q) = \tilde{\eta}(p,q) \tilde{F}^{-2}_\theta (q,p) \vv
\label{solutiong}
\end{equation}
where $\tilde{\eta}$ is a Lorentz-invariant function of $p$ and
$q$. For $\theta^{\mu \nu}=0$, $\varphi$ is a standard scalar field
and $\tilde{\eta}(p,q)$ takes the constant value $\eta = +1$. So it is
natural to take
\begin{equation}
\tilde{\eta}(p,q) = \eta = + 1, \quad {\rm for\,\ all} \,\,\theta^{\mu \nu} \vv
\end{equation}
even though $\tilde{\eta}(p,q)$ can depend on $p,q$ and $\theta^{\mu
\nu}$ and approach the value $+1$ only in the limit of vanishing
$\theta^{\mu\nu}$.

Note that in $1+1$ dimensions, $\tilde{F}_\theta\left(\Lambda p,
\Lambda q \right) = \tilde{F}_\theta\left(p,q \right)$ is itself
Lorentz-invariant (but not invariant under parity). Also, $2+1$
dimensions is special because of the availability of braid
statistics. Thus for anyons, $\tilde{\eta}(p,q)$ can be taken to be a
fixed phase.

Summarizing
\begin{equation}
a(p) a(q) = \eta \tilde{F}^{-2}_\theta\left( q,p \right) a(q) a(p) \pp
\label{newcomm2}
\end{equation}

The creation operator $a^\dagger(q)$ carries momentum $-q$, hence its
deformed relation for scalar fields is
\begin{equation}
a(p) a^\dagger(q) = \tilde{\eta}'(p,q) \tilde{F}^{-2}_\theta\left( -q,p
  \right) a^\dagger(q) a(p) + 2 p_0 \delta^d(p-q) \pp
\label{newcomm3}
\end{equation}
There is no need that $\tilde{\eta}(p,q) = \tilde{\eta}'(p,q)$, even
though as $\theta^{\mu \nu}$ approaches zero we require that
$\tilde{\eta}'(p,q)$ approaches the constant $\eta'= +1$. Hence, as
before we will set $\tilde{\eta}'(p,q) = \eta '= +1$ for all
$\theta^{\mu \nu}$.

Finally, the adjoint of (\ref{newcomm2}) gives
\begin{equation}
\bar{\eta} a^\dagger(p) a^\dagger(q) = \tilde{F}^{-2}_\theta\left( q,p
\right) a^\dagger(q) a^\dagger(p) \vv
\label{newcomm4}
\end{equation}
where $\bar{\eta}=+1$ for $\eta=+1$.

For spinorial free fields, there are similar deformed relations where
the factors $\tilde{\eta},\tilde{\eta}'$ approach $-1$ as $\theta^{\mu
\nu} \rightarrow0$.

\section{Construction of Deformed Oscillators from Undeformed
  Oscillators}

We have presented such a construction elsewhere \cite{bmt} when
considering deformations of target manifolds of fields.

Let $c(p), c^\dagger(p)$ denote the undeformed oscillators, the limits
of $a(p), a^\dagger(p)$ when $\theta^{\mu \nu} \rightarrow 0$, as in
(\ref{standcomm1}), (\ref{standcomm2}). Then
\begin{equation}
a(p) = c(p) e^{\frac{i}{2} p_\mu \theta^{\mu \nu} P_\nu} \vv
\label{aitoc}
\end{equation}
where $P_\nu$ generates translations in the Hilbert space:
\begin{equation}
P_\nu = \int \frac{d^d p}{2 p_0}\;\; p_\nu c^\dagger (p) c(p) \vv
\end{equation}
\begin{equation}
[P_\nu, a(p)] = - p_\nu a(p), \quad [P_\nu, a^\dagger(p)] = p_\nu
a^\dagger(p) \pp
\end{equation}
The adjoint of (\ref{aitoc}) also gives
\begin{equation}
a^\dagger(p) = e^{-\frac{i}{2} p_\mu \theta^{\mu \nu} P_\nu}
c^\dagger(p) \pp \label{aitoc2}
\end{equation}

Before checking that this ansatz for the $a$-oscillators works, let us
first point out that
\begin{eqnarray}
c(p) e^{\frac{i}{2} p_\mu \theta^{\mu \nu} P_\nu} &=& e^{\frac{i}{2}
  p_\mu \theta^{\mu \nu} P_\nu} c(p) \vv \\
e^{-\frac{i}{2} p_\mu \theta^{\mu \nu} P_\nu} c^\dagger(p) &=& c^\dagger (p)
e^{-\frac{i}{2} p_\mu \theta^{\mu \nu} P_\nu} \vv
\end{eqnarray}
due to the antisymmetry of $\theta^{\mu \nu}$. Hence the ordering of
factors in (\ref{aitoc}) is immaterial. Note also that
\begin{equation}
P_\nu = \int\frac{d^d p}{2p_0} p_\nu a^\dagger(p) a(p) \vv
\end{equation}
so that the map from the $c$- to the $a$-oscillators is
invertible.

We can check the relation (\ref{newcomm2}) as follows. We have
\begin{equation}
c(p) e^{\frac{i}{2} p_\mu \theta^{\mu \nu} P_\nu} c(q) e^{\frac{i}{2} q_\rho
\theta^{\rho \sigma} P_\sigma} = e^{-\frac{i}{2} p_\mu \theta^{\mu \nu}q_\nu}
c(p) c(q) e^{\frac{i}{2}(p+q)_\mu \theta^{\mu \nu} P_\nu} \pp
\end{equation}
Hence since $[c(p), c(q)]=0$ and $\theta^{\mu \nu} = -\theta^{\nu
\mu}$, we get (\ref{newcomm2}) with $\eta = +1$ for $a(p)$ defined by
(\ref{aitoc}). We can check the remaining commutation relations as
well in the same way.

\subsection{{\it{Deformed Permutation Symmetry}}}

Let ${\cal F}$ be the Fock space of the $c$-oscillators. Since the
$a$-oscillators can be constructed from the $c$'s, ${\cal F}$ is also
a representation space for the $a$-oscillators. In particular, the
Fock vacuum is annihilated by $a(p)$:
\begin{equation}
a(p) |0\rangle =0 \pp
\end{equation}
We work with the representation of the $a$-oscillators on ${\cal F}$.

Multi-particle vector states for $\theta^{\mu \nu} \neq 0$ are
obtained by applying polynomials of $a^\dagger(p)$'s on $|0\rangle$.

The number operator
\begin{equation}
N = \int \frac{d^dp}{2p_0} c^\dagger(p)c(p) \vv
\end{equation}
has the same expression in terms of $a(p)$'s and $a^\dagger(p)$'s,
\begin{equation}
N = \int \frac{d^dp}{2p_0} a^\dagger(p)a(p) \vv
\end{equation}
and has the standard commutators with these oscillators,
\begin{equation}
[N, a^\dagger(p)] = a^\dagger (p), \quad [N, a(p)] = -a(p) \pp
\end{equation}
Thus
\begin{equation}
N \prod_{i=1}^k a^\dagger (p_i)^{n_i}|0\rangle =
\left(\sum_{j=1}^k n_j \right) \left( \prod_{i=1}^k a^\dagger
(p_i)^{n_i} \right) |0 \rangle \vv
\end{equation}
and we can justifiably call
\begin{equation}
\prod_{i=1}^k (a^\dagger (p_i))^{n_i} |0 \rangle \vv
\end{equation}
as the $n$-particle state where $n = \sum_{i=1}^k n_i$.

We now show that there is a totally symmetric representation of the
permutation group on these vectors. The operator representatives of
its group elements depend on $\theta^{\mu \nu}$, but they reduce to
the standard realizations for $\theta^{\mu \nu}=0$.

First consider the free tensor product of two single particle
states. On these, we can define the transposition $\hat{\sigma}$,
\begin{equation}
\hat{\sigma} (v (p) \otimes v (q)) := v(q) \otimes v(p) \vv
\end{equation}
where
\begin{equation}
v(p) = a^\dagger (p) |0 \rangle \vv
\end{equation}
and so
\begin{equation}
\hat{\sigma}^2 = 1\!\mbox{l} \pp
\end{equation}
Here there is no relation between $v(p) \otimes v(q)$ and $v(q)
\otimes v(p)$ for a generic $v$ and all $p,q$.

The twist
\begin{equation}
\hat{F}_\theta = e^{-\frac{i}{2} P_\mu \theta^{\mu \nu}
\otimes P_\nu}
\end{equation}
acts on $v(p) \otimes v(q)$ as
\begin{equation}
\hat{F}_\theta (v (p)\otimes v (q)) = e^{-\frac{i}{2} p_\mu \theta^{\mu \nu}
  q_\nu} v (p) \otimes v (q) \pp
\end{equation}
By the antisymmetry of $\theta^{\mu \nu}$,
\begin{equation}
\hat{F}_\theta \hat{\sigma} = \hat{\sigma} \hat{F}_\theta^{-1} \vv
\end{equation}
so that
\begin{equation}
\hat{T} = \hat{F}^{-2}_\theta \hat{\sigma} \vv
\label{Tdef}
\end{equation}
is an involution:
\begin{equation}
\hat{T}^2 = 1\!\mbox{l} \pp
\end{equation}

Note that the action of neither $\hat{\sigma}$ nor $\hat{F}^{-2}_\theta$
preserves the relation (\ref{newcomm4}), while that of $\hat{T}$ does
for $\bar{\eta}=+1$. That is, if (\ref{newcomm4}) is true with
$\bar{\eta}=+1$, then so is
\begin{equation}
\hat{T} a^\dagger (p) a^\dagger (q) \hat{T}^{-1} = \tilde{F}^{-2}_\theta
(q,p) \hat{T} a^\dagger (q) a^\dagger (p) \hat{T}^{-1} \pp
\end{equation}
{\it This means that $\hat{F}^{-2}_\theta$ and $\hat{\sigma}$ individually map
the subspace ${\cal H}_S$ spanned by the vectors $\{ a^\dagger (p)
a^\dagger (q) |0\rangle \}$ out of ${\cal H}_S$ and into the full free
tensor product of two single particle subspaces, while
$\hat{F}^{-2}_\theta \hat{\sigma}$ maps ${\cal H}_S$ to ${\cal H}_S$}.

Further by (\ref{newcomm4}),
\begin{equation}
\hat{T} a^\dagger (p) a^\dagger (q) |0\rangle = a^\dagger (p)
a^\dagger (q) |0\rangle \pp
\end{equation}
For $\theta^{\mu \nu}=0$, we recover $\hat{T} = \hat{\sigma}$, the
standard representation. We therefore call $a^\dagger (p) a^\dagger
(q)|0\rangle$ as the symmetric state. Bose symmetry thus generalizes
to symmetry under $\hat{T}$.

The generalizations of $\hat{T}$ to three-particle states are the two
transpositions
\begin{equation}
\hat{T}_{12} = \hat{T} \otimes {1\!\mbox{l}}, \quad \hat{T}_{23} =
{1\!\mbox{l}} \otimes \hat{T} \pp
\end{equation}
On the $n$-particle states, $\hat{T}$ generalizes to the $(n-1)$
transpositions
\begin{equation}
\hat{T}_{i,i+1} = \underbrace{{1\!\mbox{l}} \otimes ...
\otimes}_{(i-1)\;\; {\rm factors}}  \hat{T} \otimes
\underbrace{{1\!\mbox{l}} \otimes {1\!\mbox{l}} ... \otimes
{1\!\mbox{l}}}_{n-(i+1) \;\;{\rm factors}} \pp
\end{equation}
They square to unity:
\begin{equation}
\hat{T}_{i,i+1}^2 = {1\!\mbox{l}} \pp \label{Tsquared}
\end{equation}
In addition, as one can easily verify, they fulfill the relation
\begin{equation}
\hat{T}_{i,i+1} \hat{T}_{i+1,i+2} \hat{T}_{i,i+1} =
\hat{T}_{i+1,i+2} \hat{T}_{i,i+1} \hat{T}_{i+1,i+2} \pp
\label{braid}
\end{equation}
In view of (\ref{Tsquared}) and (\ref{braid}) and a known theorem
\cite{birman}, $\hat{T}_{i,i+1}$ generate the permutation group $S_n$.

The preceding discussion shows that we get the totally symmetric
representation of $S_n$ on the physical state vectors of a scalar
field: the scalar field describes generalized bosons.

\subsection{\it{Projector for Physical States}}

Let $\hat{t}_i, \;(i=1,...,n!)$ be the representatives of the elements
of $S_n$ on ${\cal F}$. The $\hat{t}_i$ can be written in terms of
$\hat{T}_{i,i+1}$. Then, as is well-known \cite{bt},
\begin{equation}
\hat{\cal P} = \frac{1}{n!} \left( \sum_i \hat{t}_i \right) \vv
\end{equation}
is the projector to the symmetric representations of $S_n$ carried by
${\cal F}$. The physical space is
\begin{equation}
\hat{\cal P} {\cal F} \pp
\end{equation}

\subsection{\it{Observables}}

Observables $\hat{K}$ must preserve the space $\hat{\cal P}{\cal F}$:
\begin{equation}
\hat{K}\hat{\cal P}{\cal F} \subseteq \hat{\cal P}{\cal F} \pp
\end{equation}
Hence they must commute with $\hat{\cal P}$,
\begin{equation}
\hat{K} \hat{\cal P} = \hat{\cal P}\hat{K} \pp
\end{equation}
This is assured if they commute with the permutations:
\begin{eqnarray}
\hat{T}_{i,i+1} \hat{\cal P} &=& \hat{\cal P} \hat{T}_{i,i+1} \vv \\
\hat{t}_i \hat{\cal P} &=& \hat{\cal P} \hat{t}_i \pp
\end{eqnarray}

Let us check that the Poincar\'{e} transformations commute with
permutations. (They will, of course, since we arrived at the deformed
representation of permutations by requiring Poincar\'{e} invariance.)
If $U$ is the representation of the Poincar\'{e} group with elements
$g$ on the one-particle quantum states, then its representation in say
two-particle states is $(U \otimes U) \Delta_\theta$. The image of $g$
in this representation is hence
\begin{equation}
U^{(2)} (g) := \hat{F}^{-1}_\theta [U(g) \otimes U(g)]
\hat{F}_\theta \pp
\end{equation}
Now
\begin{eqnarray}
\hat{T} U^{(2)}(g) &=& \hat{F}^{-1}_\theta \hat{\sigma} [(U(g)
\otimes U(g)]\hat{F}_\theta = \hat{F}^{-1}_\theta [U(g) \otimes
U(g)] \hat{\sigma}\hat{F}_\theta \nonumber \\
&=& \hat{F}^{-1}_\theta [U(g) \otimes U(g)] \hat{F}^{-1}_\theta
\hat{\sigma} = U^{(2)} (g) \hat{T} \pp
\end{eqnarray}
This proof generalizes to the $n$-particle sectors. Hence Poincar\'{e}
transformations commute with permutations.

\section{Quantum Hall System}

Consider a particle of charge $e$, like the electron, moving in the
$1-2$-plane ${\mathbb R}^2 \subset {\mathbb R}^3$ in a constant
magnetic field $B$ directed in the third direction. The quantum
mechanical degrees of freedom can be described by two sets of mutually
commuting oscillators $a, a^\dagger$ and $b, b^\dagger$ \cite{fubini}:
\begin{equation}
[a, a^\dagger] = [b,b^\dagger]={1\!\mbox{l}} \pp
\end{equation}
All other commutators involving these operators are zero.

The Hamiltonian describing the Landau levels is
\begin{eqnarray}
H &=& \hbar \omega (a^\dagger a +1/2) \vv \\
\omega &=& \frac{e B}{m c} = {\rm cyclotron \;\; frequency} \pp
\end{eqnarray}
Thus $H$ commutes with the $b$-oscillators.

The $a$- and $b$- oscillators separately describe a Groenewold-Moyal
plane since for example
\begin{equation}
\left[\frac{b+b^\dagger}{\sqrt{2}}, \frac{b-b^\dagger}{i \sqrt{2}}
\right] = i{1\!\mbox{l}} \pp
\end{equation}
We can hence identify $(b+b^\dagger)/\sqrt{2}$ with $\hat{x}_1/l$,
$(b-b^\dagger)/(i \sqrt{2})$ with $\hat{x}_2/l$ and $\theta_{\mu \nu}$
with $l^2 \epsilon_{\mu \nu}$ ($\epsilon_{\mu \nu} = -\epsilon_{\nu
\mu}, \epsilon_{01}=+1$) where the scale factor $l$ is the magnetic
length:
\begin{equation}
l = \frac{1}{\sqrt{|e| B}} \pp
\end{equation}

The $a$-oscillators give the discrete energy levels of the charged
particle while the $b$-oscillators are associated with the coordinates
of the plane ${\mathbb R}^2$. In fact, when only the lowest Landau
levels are excited, it can be readily proved that $\hat{x}_\mu$ are
the projections of the exact spatial coordinates to the subspace
spanned by these levels. They become commuting coordinates when $B
\rightarrow \infty$. In that limit, $\omega \rightarrow \infty$ so
that the approximation of spatial coordinates by $\hat{x}_\mu$ becomes
exact. The operators $\hat{x}_\mu$ are called the ``guiding centre''
coordinates.

When just the lowest Landau level is excited, the Hilbert space is
\begin{equation}
{\cal H}_1 \otimes {\cal H}_\infty \vv
\end{equation}
where ${\cal H}_1$ has the vacuum state $|0\rangle$ of the
$a$-oscillator as basis,
\begin{equation}
a|0 \rangle =0 \vv
\end{equation}
and ${\cal H}_\infty$ is the Fock space of the $b$-oscillators. The
observables are described by the algebra ${\cal A}_\theta ({\mathbb
R}^2)$ $(\theta_{\mu \nu}=l^2 \epsilon_{\mu \nu})$ generated by
$\hat{x}_\mu$.

When $N$ Landau levels are excited, ${\cal H}_1$ becomes the
$(N+1)$-dimensional Hilbert space ${\cal H}_{N+1}$ with basis
\begin{equation}
|0\rangle \vv\ \frac{(a^\dagger)^k}{\sqrt{k!}}|0\rangle, \quad
k=1,..., N \ \pp \label{hallbasis}
\end{equation}
The $(N+1) {\times} (N+1)$ matrix algebra $Mat_{N+1}$ acts on ${\cal
H}_{N+1}$:
\begin{equation}
{\it Mat}_{N+1} {\cal H}_{N+1} \subseteq {\cal H}_{N+1} \pp
\end{equation}
The full Hilbert space is \be {\cal H}_{N+1} \otimes {\cal
H}_\infty \pp \ee

The observables are thus described by the noncommutative algebra
$Mat_{N+1} \otimes {\cal A}_\theta({\mathbb R}^2)$.

The algebra ${\cal A}_\theta({\mathbb R}^2)$ admits the action of the
diffeomorphism group ${\cal D}$ provided the coproduct for the latter
is deformed. Although the quantum Hall system is non-relativistic, we
can perhaps impose the dogma that the underlying spacetime algebra
preserves its automorphism group in the process of deformation. If we
do so, the statistics of the excitations described by
(\ref{hallbasis}) are also deformed.

We argue elsewhere \cite{bpq} that at the second-quantized level, such
excitations do not show UV-IR mixing. That is another good reason for
the adoption of the deformed coproduct and statistics.

But the physical implications of this approach remain to be
explored.

\section{Remarks on Phenomenology}

The most striking effects appear to be associated with violations of
Pauli principle, and they can be subjected to stringent experimental
tests. For example, life times for Pauli forbidden processes like
$^{16}O \rightarrow ^{16} \tilde{O}$ or $^{12}C \rightarrow ^{12}
\tilde{C}$, where $^{16} \tilde{O}$ ($^{12} \tilde{C}$) are nuclear
configurations with an extra nucleon in the (filled) $1S_{1/2}$ shell,
are presently found to be longer than $10^{27}$ years (90 \% C.L.)
\cite{sk,borexino}. Here we indicate how such transitions can arise by
studying a very simple example: that of a free (twisted) fermion
field. Spin effects are ignored as they are not important in this
context.

So let $a(p)$ and $a^\dagger(p)$ be twisted fermionic creation and
annihilation operators for momentum $p$. They can be written in the
form (\ref{aitoc}) and (\ref{aitoc2}) where $c(p)$ and its adjoint are
fermionic oscillators for $\theta^{\mu \nu}=0$. A (twisted ) single
particle wave packet state $|\alpha \rangle$ is created from the
vacuum by the operator
\begin{equation}
\langle a^\dagger, \alpha \rangle = \int \frac{d^dp}{2p_0}
\alpha(p) a^\dagger (p) \pp
\end{equation}
Thus
\begin{eqnarray}
|\alpha \rangle &=& \langle a^\dagger, \alpha \rangle |0\rangle \\
&=& \langle c^\dagger, \alpha \rangle |0\rangle \vv \\
\langle c^\dagger, \alpha \rangle &=& \int \frac{d^dp}{2p_0}
\alpha(p) c^\dagger (p) \pp
\end{eqnarray}
Hence with
\begin{equation}
\int  \frac{d^dp}{2p_0} |\alpha(p)|^2 =1 \vv
\end{equation}
$|\alpha \rangle$ is normalized to unity:
\begin{equation}
\langle \alpha | \alpha \rangle =1 \pp
\end{equation}

We can approximate a vector with sharp momentum $\vec{p}$ with
arbitrary precision with a function $\alpha$ peaked at $\vec{p}$ and
normalized to 1. A Gaussian $\alpha$ is sufficient for this purpose.

Consider next the two-particle state vector
\begin{eqnarray}
|\alpha, \alpha \rangle &=& \langle a^\dagger, \alpha \rangle
\langle a^\dagger, \alpha \rangle |0\rangle \\ &=& \int
\frac{d^dp_1}{2p_{10}}\frac{d^dp_2}{2p_{20}} e^{-\frac{i}{2} p_{1 \mu}
\theta^{\mu \nu}p_{2\nu}} \alpha(p_1) \alpha(p_2) c^\dagger (p_1)
c^\dagger (p_2) |0\rangle \pp \label{exotic2}
\end{eqnarray}
This vector is identically zero if $\theta^{\mu \nu}=0$ as required by
Pauli principle.

But this vector is not zero if $\theta^{\mu \nu} \neq 0$, as shown for
example by its non-vanishing norm $N(\alpha, \alpha)$:
\begin{eqnarray}
N^2(\alpha,\alpha) &=& \langle \alpha, \alpha|\alpha, \alpha \rangle \\
&=& \int  \frac{d^dp_1}{2p_{10}}\frac{d^dp_2}{2p_{20}}
(\bar{\alpha}(p_1) \alpha(p_1))(\bar{\alpha}(p_2)
\alpha(p_2))(1-e^{-i p_{1 \mu}\theta^{\mu \nu} p_{2\nu}}) \pp
\end{eqnarray}
$N^2(\alpha,\alpha) \neq 0$ for $\alpha \neq 0$ as can be seen from
the following argument. We have
\begin{equation}
\int \frac{d^d p_1}{2p_{10}}\frac{d^d p_2}{2p_{20}} (\bar{\alpha}(p_1)
    \alpha(p_1))(\bar{\alpha}(p_2) \alpha(p_2)) \sin (p_{1 \mu}
    \theta^{\mu \nu} p_{2 \nu}) = 0
\end{equation}
since the integrand is odd under the interchange of $p_1
\leftrightarrow p_2$. Hence
\begin{equation}
N^2 (\alpha, \alpha) = \int \frac{d^d p_1}{2p_{10}}\frac{d^d
    p_2}{2p_{20}} (\bar{\alpha}(p_1) \alpha(p_1))(\bar{\alpha}(p_2)
    \alpha(p_2)) [1-\cos(p_{1 \mu} \theta^{\mu \nu} p_{2 \nu})] \pp
\label{Nalpha}
\end{equation}
This is strictly positive for $\alpha \neq 0$ since $1-\cos(p_{1 \mu}
\theta^{\mu \nu} p_{2 \nu}) \geq0 $ for $\theta^{\mu \nu} \neq 0$ and
vanishes only on a zero-measure set of $p_1 , p_2$. Note from
(\ref{Nalpha}) that $N(\alpha,\alpha)$ is $O(\theta)$.

We can normalize $|\alpha,\alpha \rangle$:
\begin{eqnarray}
|\alpha, \alpha) &=& |\alpha,\alpha \rangle \frac{1}{N(\alpha,\alpha)} \vv \\
(\alpha,\alpha|\alpha,\alpha) &=& 1 \pp
\end{eqnarray}
This vector, being of unit norm, remains in the Hilbert space even if
$\theta^{\mu\nu} \rightarrow 0$. But the scalar product of $|\alpha,
\alpha)$ with the fermionic Fock space state $c^\dagger (p_1)
c^\dagger (p_2)|0\rangle$ is undefined in the limit $\theta^{\mu \nu}
\rightarrow 0$. Thus
\begin{equation}
\langle 0|c(p_2) c(p_1) |\alpha, \alpha) = -2i \alpha(p_1)\alpha(p_2)
\frac{\sin (p_{1\mu} \theta^{\mu \nu} p_{2\nu}/2)}{N(\alpha,\alpha)} \pp
\end{equation}
Since $N(\alpha,\alpha)$ is $O(\theta)$, the limit of this expression
as $\theta^{\mu \nu} \rightarrow 0$ depends on the manner in which
$\theta^{\mu \nu}$ goes to zero. This means that $|\alpha,\alpha)$ has
different expansions in the Fock space basis depending on the way in
which $\theta^{\mu \nu}$ becomes zero, that is it approaches different
standard fermionic vectors in the Hilbert space depending on this
limit. We do not know how to interpret this result.

Generalizing, we have the vectors
\begin{equation}
|\underbrace{\alpha, \alpha,..., \alpha}_{N \;{\rm factors}}
\rangle = \langle a^\dagger, \alpha \rangle^N |0\rangle \vv
\end{equation}
which after normalization become $|\alpha, \alpha, \ldots, \alpha)$,
$(\alpha, \ldots, \alpha | \alpha, \ldots, \alpha) =1$. These vectors
span a Hilbert space ${\cal H}_S$ of symmetric vectors when
$\theta^{\mu\nu} \rightarrow 0$.

Now consider for example
\begin{equation}
|\beta, \gamma \rangle = \langle a^\dagger , \beta \rangle \langle
a^\dagger , \gamma\rangle |0 \rangle, \quad \beta \neq \gamma.
\end{equation}
We have
\begin{equation}
\langle \beta, \gamma|\alpha, \alpha) = \int
\frac{d^dp_1}{2p_{10}}\frac{d^dp_2}{2p_{20}}(\bar{\beta}(p_1)\alpha(p_1))
(\bar{\gamma}(p_2)\alpha(p_2))[1-e^{-i p_{1 \mu}\theta^{\mu \nu}
p_{2\nu}}]\frac{1}{N(\alpha, \alpha)}\,.
\end{equation}
This overlap amplitude is not in general zero. Thus transitions are
possible between Pauli-principle allowed state vectors $|\beta, \gamma
\rangle$ and Pauli-principle forbidden state vectors
$|\alpha,\alpha)$.

It is important to note that the mean energy and momentum in these new
states are nothing outrageous. In fact, as one can see from
(\ref{exotic2}), the mean value of $P_\mu$ in
$|\alpha,\alpha,\cdots,\alpha )$ can be made arbitrarily close to $N
p_\mu$ by choosing a Gaussian for $\alpha$ which is suitably peaked at
$p_\mu$.

In conventional Fock space, by Pauli principle, there is no fermionic
state vector with energy-momentum $N p_\mu \;(N \geq 2)$ . This shows
rather clearly that Pauli principle is violated when $\theta^{\mu \nu}
\neq 0$.

We plan to further discuss the theory and phenomenology of these exotic
states and associated transitions elsewhere.

\section*{Acknowledgments}

We thank B. Qureshi for extensive discussions. The work of A.P.B. and
A.P. was supported by DOE under grant number DE-FG02-85ER40231.
A.P.B. was also supported by NSF under grant number INT 9908763 and by
CONACyT under grant number C0 141639. The visit of G.M. to Syracuse
during Spring 2005 was fully supported by Syracuse University. This
work would have been impossible without this support.

\end{document}